*Sanika Damle; Radhika Krishnan (International Institute of Information Technology, Hyderabad)*

# Framing Climate Change on YouTube: North-South Divides in Narratives and Public Engagement

## Abstract

Climate change debates unfold increasingly on social media platforms, with YouTube serving as both a news source and a space for public discourse. While prior studies have often examined climate discourse at a global level, less attention has been paid to how geopolitical divides shape narratives and public responses online. This paper presents an exploratory analysis of climate-related YouTube videos through the lens of the Global North–South divide. We analyze 758 English-language videos linked to major international climate negotiation events and their associated comment sections. Using topic modeling to examine video transcripts and sentiment analysis to study audience reactions, we identify distinct patterns in how climate issues are framed and received. Videos that originate from the Global North more frequently emphasize emission reduction policies and institutional responsibility, while those from the Global South foreground development-related concerns. Audience responses diverge more sharply: comment sections under Global North videos are dominated by criticism and conspiracy-related discourse, whereas audiences are comparatively more supportive and offer constructive arguments under Global South videos. These findings highlight a gap between curated climate narratives and public sentiment on YouTube and suggest that platform dynamics may reinforce or reshape existing geopolitical divides in climate communication.

# Introduction

Over the past several decades, climate change has emerged as one of the most urgent challenges in global governance, requiring international coordination and collective action (Awasthi Kumari & Pande, 2023). Since the establishment of the United Nations Framework Convention on Climate Change (UNFCCC) and the first Conference of the Parties (CoP) in Kyoto in 1997, international negotiations have repeatedly exposed deep-seated tensions. Central points of contention include the principle of *Common but Differentiated Responsibilities* (CBDR), the transfer of technology, and timelines for emissions reductions. These debates reflect deeper structural inequalities, primarily between industrialized nations of the Global North—such as the United States, Canada, Japan, and much of Europe—and countries of the Global South, including regions in South Asia, Africa, and Latin America (Miles, 2025; Paterson & Grubb, 1992). While the Global North has historically contributed the majority of greenhouse gas emissions, countries in the Global South have borne disproportionate consequences. Although the category itself is contested and internally diverse, the term Global South remains a useful lens for capturing evolving alliances, responsibilities, and struggles for climate justice (Bull & Banik, 2025). For the purposes of this study, Russia is treated as part of the Global South, given its recent economic trajectory and structural position in the global system align it more closely with that grouping than with the industrialized North (Grishin, 2024).

At the same time, public debates around climate change increasingly unfold on digital platforms. Among these, YouTube stands out as one of the most influential sites for political communication, reaching billions of users and serving as both a news source and a participatory arena for civic engagement. Unlike traditional media, YouTube enables both policy makers and the general public to engage in climate discourse. Moreover, the platform's recommendation systems often promote regionally proximate content, meaning that audiences are likely to engage with videos produced within similar geopolitical contexts (Brodersen et al., 2012). As a result, YouTube provides a critical site for examining how climate discourse is framed, contested, and received by the public (Munger et al., 2025).

Despite the centrality of both the geopolitical divisions in climate negotiations, and the role of digital platforms in shaping public discourse, there is limited research examining how the North–South divide manifests in platform-mediated communication. Existing computational studies of online climate discourse have primarily focused on issues such as polarization, misinformation, and emotional framing, often treating climate communication as a homogeneous global phenomenon rather than one structured by historical inequalities and geopolitical tensions (Awasthi Kumari & Pande, 2023; Bull & Banik, 2025).

This study addresses this gap by analyzing climate-related YouTube videos through the lens of the Global North–South divide. We compile a dataset of videos associated with major international climate governance events and examine both video transcripts and comment sections. By comparing content originating from Global North and Global South contexts, we investigate not only how climate issues are framed but also how audiences respond to these narratives.

Specifically, we ask:

**RQ1:** How does the framing of climate issues in YouTube videos vary between content originating from the Global North and the Global South?

**RQ2:** Do digital narratives reproduce key divides observed in international climate negotiations, such as CBDR, equity, and technology transfer?

**RQ3:** How do audience responses in YouTube comment sections vary across these regions, and what do they reveal about public positioning on these issues?

By jointly analyzing curated narratives (video content) and public reactions (comments), this study captures two interconnected layers of platform-mediated discourse. Methodologically, it integrates topic modeling (BERTopic), large language model–based stance detection, and sentiment analysis to provide a comprehensive account of both narrative construction and reception. This approach enables us to assess whether geopolitical divisions central to international climate politics are reproduced, transformed, or attenuated in online communication environments, and to better understand the role of digital platforms in shaping global climate discourse.

# Related Work

### YouTube as a Site of Climate Debate and Polarization.

Several studies have examined YouTube's role as a platform for climate change activism and civic engagement, especially during moments of heightened political attention. During the CoP15 summit, activist networks strategically used YouTube to promote climate justice campaigns, mobilize protests and coordinate direct action (Askanius & Uldam, 2011). YouTube can enable passionate and dialogic engagement around climate justice issues by providing a space for decentralized participation and alternative media practices (Uldam & Askanius, 2013). However, prior research also shows that comment sections are often characterized by declarative statements and political positioning rather than deliberative or constructive debate (Shapiro & Park, 2015).

More recent studies emphasize YouTube's role as a site of polarization and contested narratives. Platform affordances and recommendation systems can amplify emotionally engaging and controversial content, including climate skepticism and misinformation (Munger et al., 2025; Nadal, 2024). Climate-skeptical creators often post exclusionary rhetoric to cast doubt on scientific consensus and frame climate activism as elitist or disconnected from everyday concerns (Vallström & Törnberg, 2025), while comment threads frequently display impoliteness and delegitimizing discourse targeting activists (Andersson, 2021). Additionally, YouTube's algorithmic dynamics tend to privilege already popular content, which is often produced by professional media organizations in the Global North, over local or grassroots voices, shaping the visibility and reach of different narratives (Segerberg & Magnani, 2025). Collectively, these studies position YouTube as a key site for understanding how climate discourse is produced, amplified, and contested in digital environments.

## The North-South Divide in Climate Politics.

Climate politics has long been structured by tensions between developed (Global North) and developing (Global South) countries over responsibility, equity, and development priorities (Abreu Mejía, 2010). Following the Kyoto Protocol (1997), Northern countries advocated for binding emission reductions, while Southern countries emphasized their right to economic development and highlighted the North's historical emissions (Sokona & Najam, 2002). Central to these debates is the principle of c*ommon but differentiated responsibilities* (CBDR), which assigns leadership in mitigation efforts and in providing financial and technical support to the South (Voigt & Ferreira, 2016).

Subsequent negotiations have complicated this divide, as emerging economies such as China, India, and Brazil have been increasingly called upon to contribute to mitigation efforts while continuing to voice developmental concerns (Abreu Mejía, 2010). The Paris Agreement (2015) extended mitigation responsibilities to all parties while retaining the CBDR framework, and Southern nations have continued to demand climate finance and capacity-building support (Heras & Gupta, 2025). Persistent points of contention include the operationalization of CBDR, disagreements over timelines and obligations, and critiques of "eco-colonialism" in Northern environmental agendas. While these geopolitical tensions are well documented in international negotiations, little research has examined whether similar divisions manifest in digital communication spaces such as YouTube.

## Methodological Approaches to Discourse Analysis.

Recent research on climate communication has increasingly adopted computational approaches to analyze large-scale social media data. Topic modeling techniques, particularly transformer-based methods such as BERTopic, have been widely used to uncover latent themes in climate discourse. These methods demonstrate improved coherence and interpretability compared to traditional methods like Latent Dirichlet Allocation (LDA) (Gokcimen & Das, 2024; Grootendorst, 2022). Comparative studies further show that BERTopic is effective in capturing temporal shifts in discourse, outperforming dynamic topic models in certain contexts (Adam & Kogler, 2023).

Beyond topic modeling, sentiment analysis and emotion detection have been used to examine public responses to climate content, revealing how affective dimensions shape engagement and interpretation (Meza et al., 2018). More recent work extends these approaches using machine-in-the-loop and large language model (LLM)–assisted methods for stance detection, argument mining, and narrative analysis. These enable scalable analysis of complex discourse while introducing challenges related to evaluation and interpretability (Islam & Goldwasser, 2025a, 2025b; Islam et al., 2023, 2025). Cross-platform studies highlight that discourse patterns vary significantly across platforms such as YouTube, TikTok, and Twitter, highlighting the importance of platform-specific analytical strategies (Pera & Aiello, 2024).

Recently, work that integrates multiple techniques, like combining topic modeling, sentiment analysis and network-based approaches has grown in popularity, which are able to produce more comprehensive accounts of online discourse (Molenaar et al., 2024; Domínguez et al., 2024). While these methods provide powerful tools for analyzing climate communication at scale, most studies adopt a global or platform-wide perspective, often overlooking how discourse is shaped by underlying geopolitical inequalities.

### Research Gap.

While prior work provides valuable insights into the dynamics of online climate discourse, it largely treats such discourse as a homogeneous global phenomenon. In contrast, the geopolitical divisions that structure international climate negotiations—particularly the Global North–South divide—remain underexplored in platform-mediated contexts. This study addresses this gap by examining how these divisions are reflected in both content production and audience reception on YouTube, offering a comparative analysis of climate narratives across Global North and Global South contexts during prominent political moments.

## Methodology

### Data Collection.

This study examines YouTube discourse around seven pivotal moments in global climate governance that foreground tensions between the Global North and Global South: the Kyoto Protocol (1997), COP13 (CBDR, 2007), COP15 (Copenhagen Accord, 2009), the Green Climate Fund (2010), the Paris Agreement (2015), the UN Climate Action Summit (2019), and COP28 (2023). Table 1 has an overview of the selection criteria and the primary issues of debate associated with each of these moments. These events were deliberately selected to capture instances where questions of responsibility, finance, and mitigation obligations were explicitly contested, rather than relying on generic climate-related keywords that may retrieve diffuse or weakly relevant discussions. We initially also considered a temporal analysis of the evolution of climate change debates, but since YouTube's popularity has surged mainly in the past decade, the majority of available videos were concentrated in this period, making such an approach unfeasible.

| Year | Event | Key Points |
|---|---|---|
| 1997 | Kyoto Protocol | First binding emission targets for developed nations; acknowledged historical responsibility (Satish, 2024). |
| 2007 | COP13 – CBDR Principle | Formalized “common but differentiated responsibilities”; developed nations to lead mitigation (United Nations Framework Convention on Climate Change, 2008). |
| 2009 | COP15 – Copenhagen Accord | Non-binding deal; pledged $100B yearly for developing nations (United Nations Framework Convention on Climate Change, 2009). |
| 2010 | Green Climate Fund | Created climate finance mechanism; Global North fell short of $100B goal (Thwaites & Guy, 2023). |
| 2015 | Paris Agreement | Introduced NDCs; developing nations pushed for tech transfer and financial support (United Nations Framework Convention on Climate Change, 2015). |
| 2019 | UN Climate Action Summit | South criticized lack of finance; Greta Thunberg’s speech symbolized urgency (Rosane, 2021). |
| 2023 | COP28 | Agreed fossil fuel phase-out and loss-and-damage fund; South pushed for stronger financial commitments (Morton et al., 2023). |

**Table 1: Key moments in global climate governance highlighting North–South dynamics**

For each event, five targeted search queries were constructed (e.g., variations of event names, outcomes, and key highlights). Using the YouTube Data API v3, up to 400 of the most relevant videos per event were retrieved (posted up to November 2024), along with associated metadata and comments. The dataset includes content from a diverse range of sources, including news organizations, institutional channels, advocacy groups, and individual creators, reflecting the diverse nature of climate communication on the platform.

English-language transcripts were collected using the *youtube-transcript-api*. After removing duplicates and filtering for English-language content, the final dataset consisted of 758 videos. Videos were categorized as originating from the Global North or Global South based on the geographic location of the uploading channel.

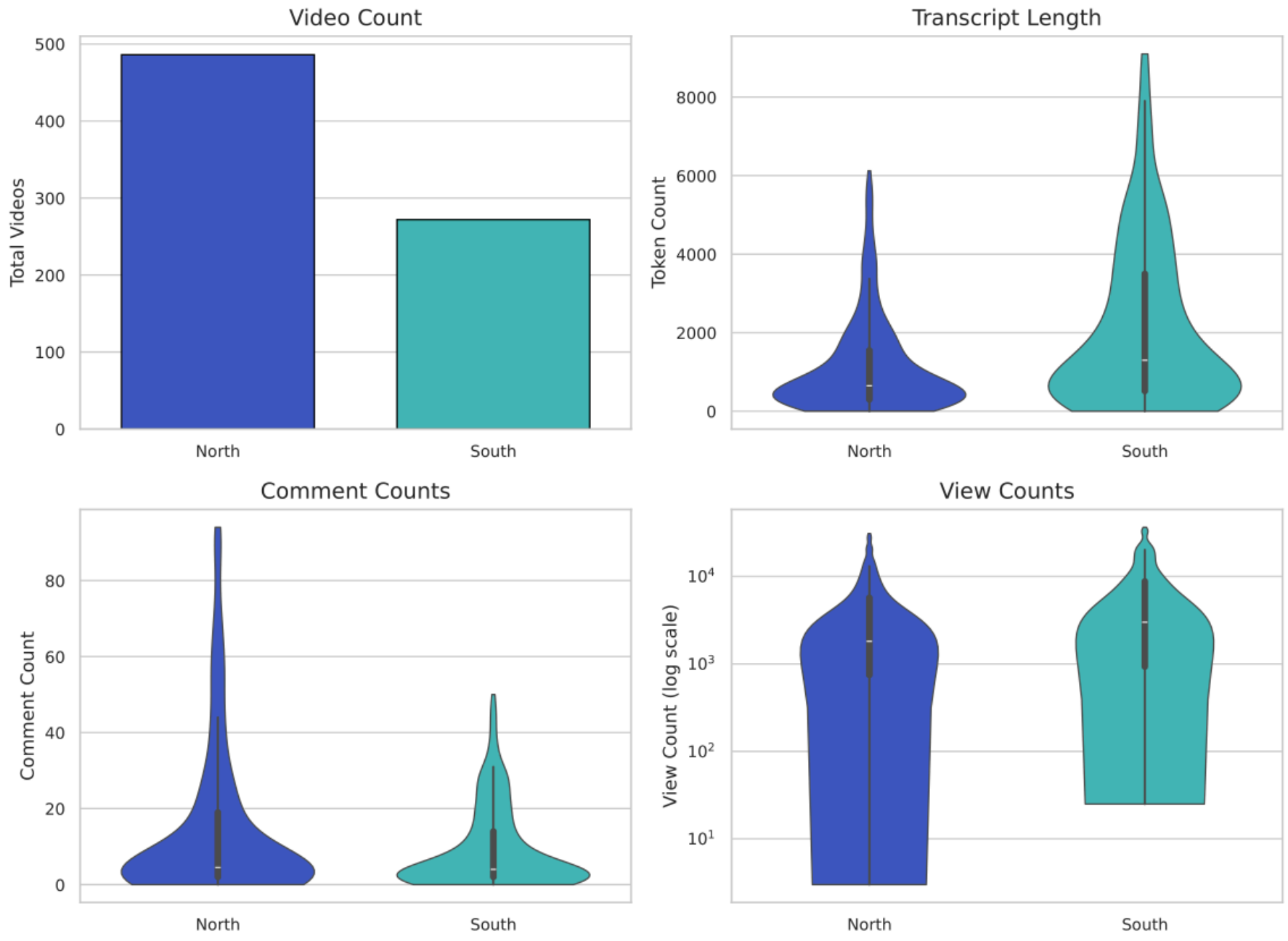

**Figure 1: Dataset composition by region and content type.** Overview of the climate-related YouTube dataset analyzed in this study, showing the distribution of videos, total comments, and average engagement metrics across Global North and Global South sources.

Although the dataset contained more videos from the Global North, no balancing was applied. Engagement-related metrics such as transcript length, number of comments, and view counts were comparable across regions. Preserving this natural distribution was done to avoid reducing the diversity of discourse captured in these videos and also reflects the content landscape encountered by users on the platform.

## Topic Extraction.

To identify dominant narratives, we applied BERTopic separately to Global North and Global South video transcripts. Specifically, we employed the KeyBERT-inspired variant, which leverages c-TF-IDF representations to generate semantically coherent clusters.

Initial topic modeling produced a larger number of clusters (13 for the Global North and 9 for the Global South). These were manually inspected for coherence and interpretability by examining top keywords and representative transcripts. Noisy or redundant clusters were removed, and the remaining topics were combined into six high-level topics per region for comparative analysis.

A similar procedure was applied to video comments. To prevent over-representation from highly viral videos, a maximum of 1000 comments per video was included. Topic modeling on comments resulted in eight broad themes capturing audience reactions, discourse patterns, and framing dynamics.

Topic modeling is used here as an exploratory method to surface recurring structures in discourse. Model outputs were therefore manually reviewed and consolidated into higher-level thematic frames based on semantic similarity and interpretability.

### Stance and Framing Signals (Supporting Analysis).

To complement the topic-level analysis, we extract coarse-grained framing signals from video transcripts using a large language model (Gemini-2.5-Flash). This analysis is intended as a supporting, aggregate-level signal, rather than a definitive classification of individual videos.

For each transcript, an instruction-guided prompting framework was used to estimate the extent to which each piece of text:

1) affirms the scientific consensus on climate change,
2) frames emissions reduction as urgent, and
3) attributes primary mitigation responsibility to actors in the Global North or South..

Each dimension was scored on a continuous scale from 0 to 1. To reduce stochastic variability, scores were computed across three independent runs and averaged. The resulting signals were aggregated at the regional level to identify broad differences in framing patterns.

Importantly, these measures are not interpreted as precise or objective classifications. Instead, they provide a coarse-grained, model-assisted lens on framing tendencies, which is further validated through manual inspection of a random sample of transcripts to ensure qualitative alignment.

### Sentiment Analysis.

Audience sentiment was analyzed using the *cardiffnlp/twitter-roberta-base-sentiment-latest* model applied to video comments. This model was selected based on manual inspection, which indicated better handling of short, informal, and sarcastic comments compared to alternative approaches. For each video's comment section, sentiment distributions across three categories—negative, neutral, and positive—were computed and aggregated. These distributions were used to compare audience responses across Global North and Global South content.

As with other automated methods used in this study, sentiment results are interpreted at an aggregate level. Comments are often brief and context-dependent, and may not directly engage with climate-related content, limiting the precision of fine-grained interpretation.

### Limitations

These analyses rely on platform-derived data and automated methods, and are therefore subject to several limitations. YouTube's recommendation and ranking systems, language filtering, and keyword-based retrieval may introduce sampling biases that cannot be fully controlled. Additionally, automated topic modeling, sentiment analysis, and LLM-based framing assessments are inherently approximate.

Accordingly, the findings are intended to capture dominant, aggregate-level patterns in climate discourse rather than provide exhaustive or individual-level representations.

# Results and Discussion

## RQ1. How does the framing of climate issues in YouTube videos vary between content originating from the Global North and the Global South?

Topic modeling of video transcripts yielded six dominant themes per region (see Table 2). These topics capture recurring patterns in how climate issues are framed across Global North and Global South content. For interpretive clarity, topics were further classified as *convergent* or *divergent* based on the degree of alignment in framing across regions.

Across both regions, a clear convergence emerges around two central themes: **emissions reduction** and **international climate agreements**. Videos from both the Global North and Global South emphasize the importance of reducing greenhouse gas emissions and reference key global accords such as Kyoto, Paris, and Copenhagen. This suggests a shared recognition of climate change as a global problem structured by international cooperation and institutional frameworks.

However, beyond these shared reference points, framing diverges in systematic ways. Global North content places stronger emphasis on **policy instruments, technological innovation, and energy transitions**, particularly the phasing out of fossil fuels and the development of renewable energy systems. Discussions in this context are often embedded within broader narratives of governance, leadership, and coordinated global action. In contrast, Global South content foregrounds **developmental priorities and localized impacts**. Prominent themes include agriculture, biodiversity, and the protection of natural systems, reflecting the region's structural reliance on climate-sensitive sectors and ecosystem services. Additionally, Global South narratives emphasize **collective action and future pathways**, often framed in terms of balancing climate mitigation with economic growth and development needs.

| Broad Topic | Global North Emphasis | Global South Emphasis | Framing Pattern |
|---|---|---|---|
| Emissions & Greenhouse Gases | Strong focus on mitigation targets and carbon reduction commitments | Present but linked to development constraints | Convergent (different emphasis) |
| Climate Agreements & Global Negotiations | Emphasis on Paris/Kyoto accords, compliance and leadership | Focus on negotiation fairness and the role of developing nations | Convergent |
| Finance, Equity & Responsibility | Climate finance framed around accountability and burden-sharing | Central focus on funding access, equity and redistribution | Divergent |
| Energy Transition & Fossil Fuels | Emphasis on phasing out coal, oil, and gas | Focus on energy access, transition costs, and economic growth | Divergent |
| Scientific Evidence & Future Risks | Climate science, projections, and global risk narratives | Humanitarian impacts, regional vulnerability, and adaptation | Divergent |
| Collective Action & Future Pathways | Institutional leadership and global coordination | Community-level action and development-oriented futures | Divergent |

**Table 2: Comparison of Dominant Climate Framing in Global North and Global South YouTube Videos**

While both regions engage with questions of **finance, equity, and responsibility**, the emphasis differs. In the Global North, these discussions tend to focus on mechanisms of funding, burden-sharing, and institutional accountability. In the Global South, they are more closely tied to **access to resources, development constraints, and unequal economic impacts**, highlighting structural asymmetries in climate governance.

Taken together, these findings suggest that while climate discourse on YouTube draws on a shared set of global policy narratives, regional framings reflect distinct socio-economic contexts and priorities. Rather than representing opposing viewpoints, these differences highlight complementary but unevenly distributed dimensions of climate politics.

### RQ2. Do digital narratives reproduce key divides observed in international climate negotiations, such as CBDR, equity, and technology transfer?

To further examine how responsibility is framed within these narratives, we analyze aggregated stance signals derived from LLM-assisted assessment (Figure 2). These signals are interpreted as coarse-grained, comparative indicators, rather than definitive classifications. Across both regions, videos show consistently high support for **climate change acceptance** and the **urgency of emissions reduction**, indicating broad alignment on the existence and importance of the issue. However, differences emerge in how responsibility for mitigation is framed.

Global North videos exhibit stronger emphasis on **collective responsibility and leadership in emissions reduction**, often aligning with narratives of coordinated global action and policy-driven mitigation. In contrast, Global South videos display comparatively lower support for assigning primary responsibility to developing regions, instead reflecting a more differentiated view of obligations. These patterns resonate with longstanding debates in international climate politics, particularly around **common but differentiated responsibilities (CBDR), equity, and resource distribution**. While Northern discourse emphasizes financial transfers, innovation, and institutional reform, Southern discourse highlights development needs, vulnerability, and the unequal burden of climate action.

Importantly, these findings should be interpreted as indicative rather than conclusive. The stance signals are derived from automated, prompt-based evaluations and are used only at an aggregate level to highlight broad framing tendencies. Nevertheless, the alignment between these patterns and established geopolitical divides suggests that digital climate narratives on YouTube partially reproduce the fault lines observed in formal climate negotiations.

At the same time, it is important to note that YouTube videos represent a relatively curated form of discourse, often produced by institutional actors, media organizations, or informed creators. As such, they may reflect a narrower subset of perspectives. To better understand how these narratives are received and interpreted by broader audiences, we turn to the analysis of comment sections.

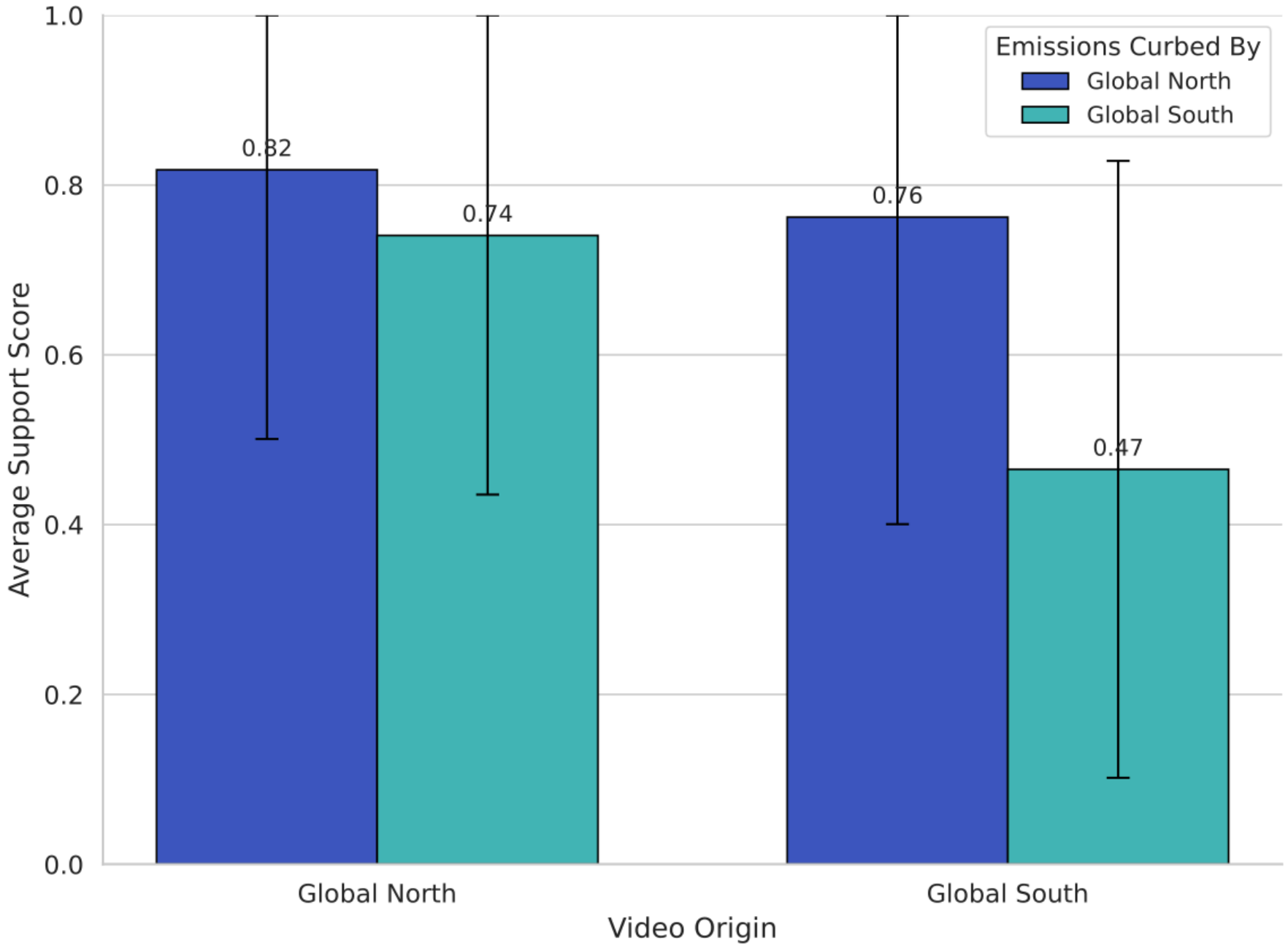

**Figure 2: Average framing-based support for emission reduction responsibility by video origin.** Aggregate scores reflect the extent to which climate-related videos frame emissions reduction as the responsibility of actors in the Global North or Global South. Scores range from 0 to 1 and are used only for comparative, region-level analysis

## RQ3. How do audience responses in YouTube comment sections vary across these regions, and what do they reveal about public positioning on these issues?

Topic modeling of video comments (see Tables 3 and 4) reveals distinct patterns in how audiences engage with climate-related content across regions. Comment-level topics were grouped into broad categories such as *discussion*, *criticism*, *support*, *conspiracy*, and *agitation* based on manual inspection.

In the Global North, comment sections are characterized by a high prevalence of **criticism, conspiracy, and oppositional discourse**, alongside general discussion. A substantial portion of comments reflects skepticism, distrust, or mockery, indicating a more polarized and confrontational engagement environment. While supportive and constructive comments are present, they are comparatively less frequent. Some discussions also express **hopelessness or fatalism**, including references to external or non-scientific solutions.

In contrast, comment sections under Global South videos exhibit a more **balanced and constructive tone**. Supportive and appreciative comments are more prominent, often recognizing the informational or educational value of the content. Discussion-oriented engagement remains prevalent, while instances of conspiracy and overt criticism are less frequent. Overall, the discourse appears less polarized and more oriented toward knowledge-sharing and engagement.

| Approx #Comments | Broad Topic | Label |
|---|---|---|
| 3,000 | Climate Science & Warming | Discussion |
| 2,500 | Politics & Leadership | Criticism |
| 2,000 | Conspiracy & Denialism | Conspiracy |
| 1,600 | Fossil Fuels & Energy | Discussion |
| 1,300 | Hypocrisy & Blame | Criticism |
| 1,200 | Appreciation & Praise | Support |
| 1,00 | Insults & Mockery | Criticism |
| 800 | Hope & Urgency | Agitation |
| 400 | Religion & Philosophy | Discussion |
| 700 | Humor & Memes | Criticism |

**Table 3: Results of the topic modeling for the Global North comments.** 20 topics were condensed down to 10 based on contextual similarity.

| Approx #Comments | Broad Topic | Label |
|---|---|---|
| 1,300 | Appreciation & Praise | Support |
| 850 | Climate Science & Environment | Discussion |
| 450 | Politics & Governance | Discussion |
| 400 | Energy & Solutions | Discussion |
| 400 | Requests & Interactions | Discussion |
| 300 | General Support & Agreement | Support |
| 300 | Hope & Urgency | Agitation |
| 200 | Conspiracy & Distrust | Conspiracy |
| 250 | National Identity & Geopolitics | Discussion |
| 200 | Insults & Mockery | Criticism |

**Table 4: Results of the topic modeling for the Global South comments.** 15 topics were condensed down to 10 based on contextual similarity.

These patterns are reinforced by sentiment analysis (Figure 3). A majority of Global North videos (63.8%) are associated with predominantly negative comment sections, with relatively smaller proportions of neutral and positive engagement. By comparison, Global

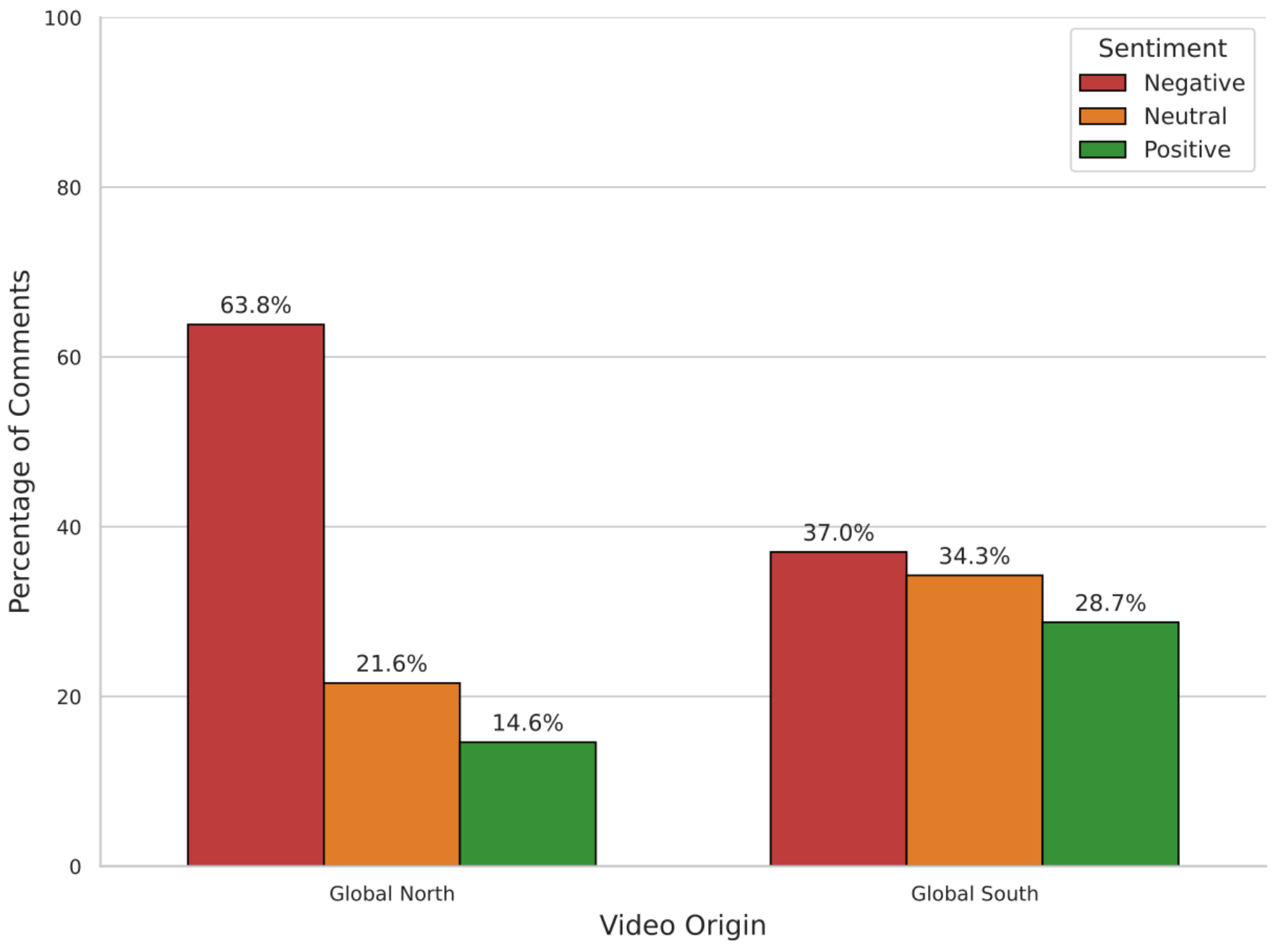

**Figure 3: Predominant sentiment in comment sections by video origin.** Proportion of climate-related videos categorized by the dominant sentiment of their associated comment sections (mode of comment-level sentiment per video), comparing Global North and Global South content.

South videos display a more even distribution of sentiment, with higher shares of neutral and positive responses.

Further analysis examining sentiment by video stance (Figure 4) reveals that, for Global North content, negative sentiment persists regardless of how responsibility is framed. Whether a video attributes emissions reduction responsibility to the North or the South, audience responses remain largely critical. This suggests that polarization in these contexts may operate independently of specific narrative positions, reflecting broader patterns of political disillusionment or climate fatigue.

For Global South content, a more nuanced pattern emerges. Videos that emphasize greater responsibility for the Global North tend to attract relatively higher negative engagement compared to those that distribute responsibility more broadly. This may indicate increasing audience sensitivity to framing responsibility in binary terms, and a growing recognition that effective climate action requires shared but differentiated accountability.

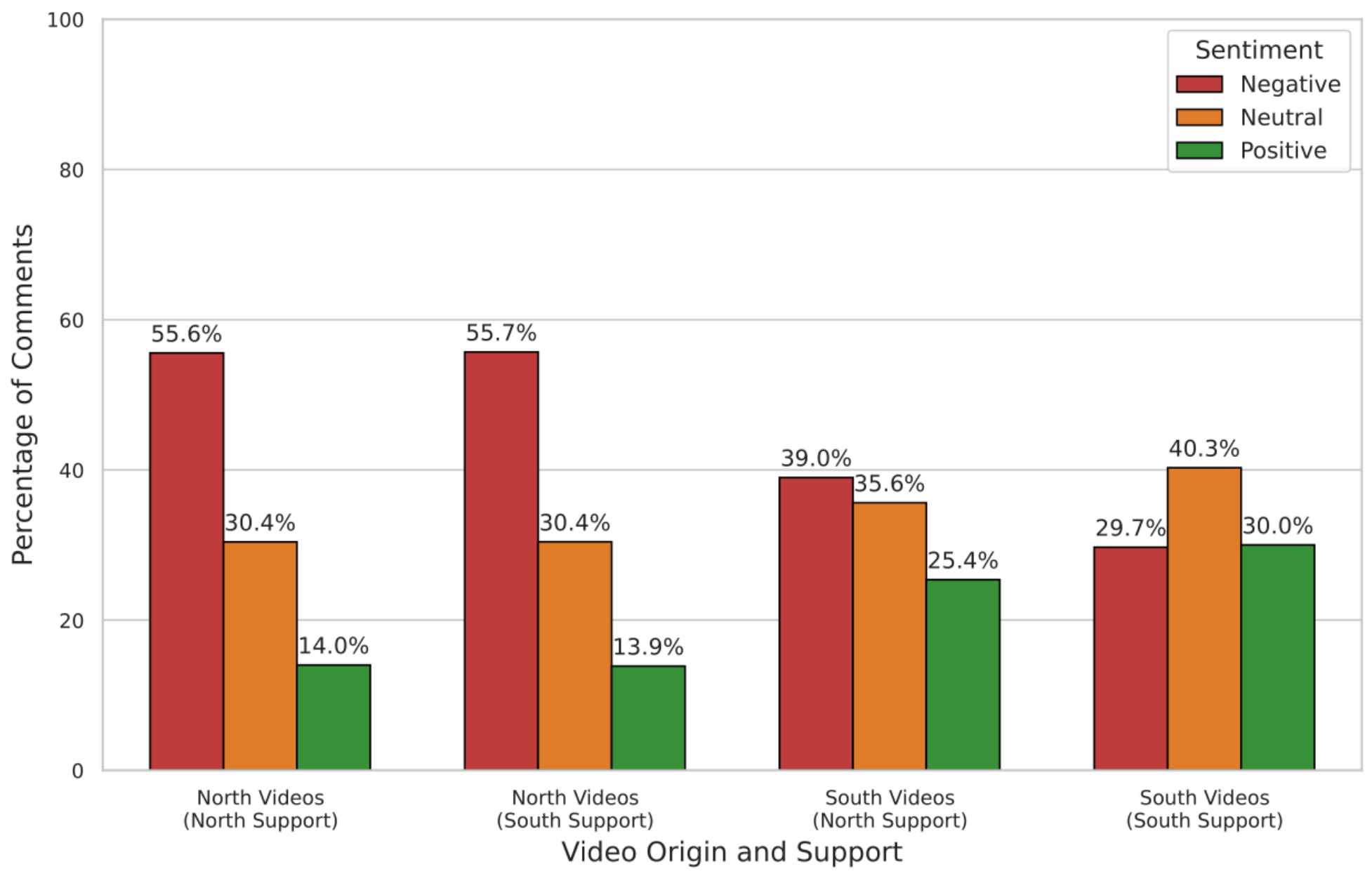

**Figure 4: Sentiment Distribution of Comments by Video Origin and Support to Curb Emissions.** Same metric as Figure 3, but divided based on the stance of the video. North Videos (North Support) indicates videos posted by the Global North that have scores higher than 0.5 in support of the North needing to cut down on emissions.

Taken together, these findings highlight a notable divergence between curated narratives and public engagement. While video content across regions reflects structured policy debates and institutional framings, audience responses, particularly in the Global North are shaped by affective reactions, skepticism, and polarization. In contrast, Global South audiences appear more receptive to informational content and constructive dialogue, potentially reflecting differing socio-political contexts and levels of direct exposure to climate impacts.

## Discussion and Implications

Across all three research questions, a consistent pattern emerges: climate discourse on YouTube reflects both convergence at the level of global narratives and divergence in framing, interpretation, and public engagement.

At the level of content production, both regions draw on shared institutional frameworks and policy debates, but adapt them to local priorities and constraints. At the level of audience response, however, differences become more apparent, with Global North discourse exhibiting higher polarization and Global South discourse showing more balanced engagement. These findings underscore YouTube's dual role as both an informational platform and a space for the expression of broader societal tensions. They also point to the need to better connect expert-driven narratives with how people actually understand and respond to them, especially in situations where polarization can make it harder to build consensus.

Finally, these results should be interpreted as exploratory and context-dependent. The analyses rely on automated methods and platform-derived data, and reflect patterns observed during specific moments of heightened attention to climate policy. Accordingly, they are best understood as indicative of dominant trends rather than exhaustive representations of global climate discourse.

# References

1. Abreu Mejía, D. (2010). The evolution of the climate change regime: Beyond a North–South divide? Institut Català Internacional per la Pau.
2. Adam, R., & Kogler, M. (2023). Tracking the evolution of climate discourse using BERTopic. In Scientific Computing 2023 (pp. 208–217).
3. Andersson, M. (2021). The climate of climate change: Impoliteness as a hallmark of homophily in YouTube comment threads. Journal of Pragmatics. https://doi.org/10.1016/j.pragma.2021.03.003
4. Askanius, T., & Uldam, J. (2011). Online social media for radical politics. International Journal of Electronic Governance, 4, 69–84.
5. Awasthi Kumari, K., & Pande, R. (2023, June 29). The imperative of collective action in a fragmented world. World Economic Forum. https://www.weforum.org/stories/2023/06/uniting-for-change-the-imperative-of-collective-action-in-a-fragmented-world/
6. Brodersen, A., Scellato, S., & Wattenhofer, M. (2012). YouTube around the world. In WWW '12 (pp. 241–250). ACM.
7. Bull, B., & Banik, D. (2025). The rebirth of the Global South: Geopolitics, imageries and developmental realities. Forum for Development Studies, 52(2), 195–214. https://doi.org/10.1080/08039410.2025.2490696
8. Domínguez, A. G., Roig-Tierno, N., Chaparro-Banegas, N., & García-Álvarez-Coque, J.-M. (2024). NLP for evaluating policy. Journal of Rural Studies, 109, 103341.
9. Gokcimen, T., & Das, B. (2024). Exploring climate change discourse using topic modeling. Heliyon, 10(11), e32464.
10. Grishin, V. (2024, March 14). Russia and the Global South, or the mystery of political semantics. https://therussiaprogram.org/gs_mystery
11. Grootendorst, M. (2022). BERTopic: Neural topic modeling with a class-based TF-IDF procedure. arXiv.

12. Heras, A., & Gupta, J. (2025). North–south relations, responsibilities, and agendas in Earth system governance: Have these changed in the Anthropocene? Earth System Governance, 24, 100251. https://doi.org/10.1016/j.esg.2025.100251
13. Islam, T., & Goldwasser, D. (2025). Post-hoc study of climate microtargeting on social media ads with LLMs: Thematic insights and fairness evaluation. In Findings of the Association for Computational Linguistics: EMNLP 2025 (pp. 15838–15859). Association for Computational Linguistics. https://doi.org/10.18653/v1/2025.findings-emnlp.857
14. Islam, T., & Goldwasser, D. (2025a). Discovering latent themes in social media messaging: A machine-in-the-loop approach integrating LLMs. Proceedings of the International AAAI Conference on Web and Social Media, 19(1), 859–884. https://doi.org/10.1609/icwsm.v19i1.35850
15. Islam, T., & Goldwasser, D. (2025b). Uncovering latent arguments in social media messaging by employing LLMs-in-the-loop strategy. In Findings of the Association for Computational Linguistics: NAACL 2025 (pp. 7397–7429). Association for Computational Linguistics. https://doi.org/10.18653/v1/2025.findings-naacl.413
16. Islam, T., Zhang, R., & Goldwasser, D. (2023). Analysis of climate campaigns on social media using Bayesian model averaging. In Proceedings of the AAAI/ACM Conference on AI, Ethics, and Society (AIES '23) (pp. 15–25). ACM. https://doi.org/10.1145/3600211.3604665
17. Meza, X., Shapiro, M., & Park, H. (2018). Climate change emotions on YouTube. Korean Data Analysis Society, 20, 1697–1708.
18. Miles, K. (2025, February 21). Global North and Global South. Encyclopedia Britannica. https://www.britannica.com/topic/Global-North-and-Global-South
19. Molenaar, A., Lukose, D., Brennan, L., Jenkins, E., & McCaffrey, T. (2024). NLP for food security discourse. Journal of Medical Internet Research, 26, e47826.
20. Morton, A., Greenfield, P., Harvey, F., Lakhani, N., & Carrington, D. (2023, December). Cop28 landmark deal agreed to 'transition away' from fossil fuels. The Guardian.
21. Munger, K., Bisbee, J., Yalcin, O., Phillips, J., & Hindman, M. (2025). Pressing play on politics: Quantitative description of YouTube. Journal of Quantitative Description: Digital Media, 5, 1–37.
22. Nadal, L. de. (2024). From denial to the culture wars: A study of climate misinformation on YouTube. Environmental Communication, 18, 1186–1203. https://doi.org/10.1080/17524032.2024.2363861
23. Paterson, M., & Grubb, M. (1992). The international politics of climate change. International Affairs, 68(2), 293–310.
24. Pera, A., & Aiello, L. M. (2024). Shifting climates: Climate change communication from YouTube to TikTok. In Proceedings of WebSci '24 (pp. 1–6). ACM.
25. Rosane, O. (2021, December). What to expect from today's UN climate action summit. EcoWatch.
26. Satish, S. (2024, November). Kyoto Protocol, 1997. ClearIAS.

27. Segerberg, A., & Magnani, M. (2025). Visual digital intermediaries and global climate communication. PLOS ONE, 20(4), e0318338.
28. Shapiro, M. A., & Park, H. (2015). More than entertainment: YouTube and public responses to climate change. Social Science Information, 54, 115–145.
29. Sokona, Y., & Najam, A. (2002). Climate change and sustainable development: Views from the South.
30. Thwaites, J., & Guy, B. (2023, April 20). U.S. delivers for the Green Climate Fund. NRDC.
31. Uldam, J., & Askanius, T. (2013). Online civic cultures? International Journal of Communication, 7, 1–20.
32. United Nations Framework Convention on Climate Change. (2008). Report of the Conference of the Parties (CBDR).
33. United Nations Framework Convention on Climate Change. (2009). Copenhagen Accord (COP15).
34. United Nations Framework Convention on Climate Change. (2015). The Paris Agreement. https://unfccc.int/process-and-meetings/the-paris-agreement
35. Vallström, V., & Törnberg, A. (2025). From YouTube to parliament: The dual role of political influencers in shaping climate change discourse. Environmental Sociology. https://doi.org/10.1080/23251042.2025.2475519
36. Voigt, C., & Ferreira, F. (2016). “Dynamic differentiation”: The principles of CBDR-RC, progression and highest possible ambition in the Paris Agreement. Transnational Environmental Law, 5(2), 285–303. https://doi.org/10.1017/S2047102516000212